\documentclass[a4paper]{jpconf}
\usepackage{graphicx}
\usepackage{amsmath}
\usepackage{amssymb}

\newcommand{\dd}{\partial}

\begin{document}
\title{Beam energy scan using a viscous hydro+cascade model}

\author{Iu.A.~Karpenko$^{1,2}$, M. Bleicher$^{1,3}$, P. Huovinen$^{1}$ and H. Petersen$^{1}$}

\address{$^1$ Frankfurt Institute for Advanced Studies, Ruth-Moufang-Stra{\ss}e 1, 60438 Frankfurt am Main, Germany}
\address{$^2$ Bogolyubov Institute for Theoretical Physics, 14-b, Metrolohichna str., 03680 Kiev, Ukraine}
\address{$^3$ Institute for Theoretical Physics, Johann Wolfgang Goethe Universit\"at, Max-von-Laue-Str. 1, 60438 Frankfurt am Main, Germany}
\ead{karpenko@fias.uni-frankfurt.de}

\begin{abstract}
 Following the experimental program at BNL RHIC, we perform a similar ``energy scan'' using 3+1D viscous hydrodynamics coupled to the UrQMD hadron cascade, and study the collision energy dependence of pion and kaon rapidity distributions and $m_T$-spectra, as well as charged hadron elliptic flow. To this aim the equation of state for finite baryon density from a Chiral model coupled to the Polyakov loop is employed for hydrodynamic stage. 3D initial conditions from UrQMD are used to study gradual deviation from boost-invariant scaling flow.
We find that the inclusion of shear viscosity in the hydrodynamic stage of evolution consistently improves the description of the data for Pb-Pb collisions at CERN SPS, as well as of the elliptic flow measurements for Au-Au collisions in the Beam Energy Scan (BES) program at BNL RHIC. The suggested value of shear viscosity is $\eta/s\ge0.2$ for $\sqrt{s_{NN}}=6.3\dots39$~A~GeV.
\end{abstract}

\section{Introduction}
During the recent years there has been considerable progress in modelling of bulk matter dynamics in ultrarelativistic heavy ion collisions at BNL Relativistic Heavy Ion Collider (RHIC) and CERN Large Hadron Collider (LHC). The state-of-the-art theoretical descriptions are based on viscous hydrodynamics coupled to a hadron cascade. They successfully describe bulk observables such as radial flow, elliptic and higher order flow harmonics \cite{huovinenReview}, femtoscopy \cite{Karp} etc.
The idea of the present study is to apply a hybrid (or hydro+cascade) model to heavy ion collisions at lower collision energies to study how well the model can describe the existing data from CERN Super Proton Synchrotron (SPS) and recent results from Beam Energy Scan program at RHIC.

\section{Model description}
The lower the collision energy, the worse the approximation of boost-invariant scaling flow is. Thus it is important to employ initial conditions with non-trivial rapidity dependence. We use the Ultrarelativistic Quantum Molecular Dynamics (UrQMD) model \cite{urqmd} for the description of initial stage dynamics. The two nuclei are initialized according to Wood-Saxon distributions and the binary interactions are taken into account until a hypersurface at constant $\tau=\sqrt{t^2-z^2}$. We average over many UrQMD events to get smooth distribution of particles. The energy and momentum of particles is then converted to energy and momentum densities of the fluid. In addition to energy/momentum densities, initial baryon and charge densities are non-zero and obtained 
from UrQMD, evolved in the hydro stage and accounted for in the particlization procedure. On the other hand, net strangeness density is set to zero.

The hydrodynamic equations are solved in Milne ($\tau-\eta$) coordinates. The starting time of hydrodynamic evolution is chosen to be
 $\tau_0=2R/\sqrt{(\sqrt{s}/2m_N)^2-1}\label{tstart}$, where $R$ is a radius of nucleus and $m_N$ is a nucleon mass. This corresponds to the time when the two nuclei have passed through each other.

For beam energy scan, an equation of state (EoS) for finite baryon density must be used. We employ the Chiral model based EoS \cite{ChiralEoS}, which features correct asymptotic degrees of freedom, i.e.\ quarks and gluons at high temperature and hadrons in the low-temperature limits, crossover-type transition between confined and deconfined matter for all values of $\mu_B$ and qualitatively agrees with lattice QCD data at $\mu_B=0$.

The hydrodynamic code used solves the equations of relativistic viscous hydrodynamics in Israel-Stewart framework \cite{IsraelStewart}. In particular we solve the following equations for the shear stress tensor:
\begin{equation}
\langle u^\gamma \dd_{;\gamma} \pi^{\mu\nu}\rangle=-\frac{\pi^{\mu\nu}-\pi_\text{NS}^{\mu\nu}}{\tau_\pi}-\frac 4 3 \pi^{\mu\nu}\dd_{;\gamma}u^\gamma
\end{equation}
where $\dd_{;\gamma}$ denotes covariant derivative and the brackets $\langle A^{\mu\nu}\rangle$ denote the symmetric, traceless and orthogonal to $u^\mu$ part of $A^{\mu\nu}$. For the purpose of the current study we consider only the effects of shear viscosity, fixing bulk viscosity to zero, $\zeta/s=0$. We do not include the baryon/electric charge diffusion either.
For viscous hydro simulations, we initialize the shear stress tensor to zero. The relaxation time for shear, $\tau_\pi$, is taken as $\tau_\pi=3\eta/(sT)$.

The transition from fluid to particle description (so-called particlization) is made at the constant energy density $\epsilon_\text{sw}=0.5$~GeV/fm$^3$ when the medium has already hadronized (as in previous studies at $\sqrt{s_{NN}}=200$ A GeV RHIC and LHC energies \cite{Karp}). Cornelius subroutine \cite{Cornelius} is used to calculate the 3-volume elements $d\sigma_\mu$ of the transition hypersurface. It is important to note that while we use fixed energy density as transition criterion for all collision energies, the value of net baryon density on this surface is non-uniform and its average increases with decreasing collision energy. This corresponds to the increase of average temperature and decrease of baryon chemical potential with increasing collision energy, resembling the results for the collision energy dependence of chemical freezeout parameters from thermal model studies \cite{thermalModel}.
Also, since Chiral EoS deviates from free hadron-resonance gas (HRG) in hadronic phase, we switch to free HRG EoS when sampling the particles according to Cooper-Frye prescription. We recalculate the energy density, pressure, flow velocity and corresponding thermodynamical quantities from energy-momentum tensor using free HRG EoS, and employ them when sampling. Finally, we use the same corrections to the local equilibrium distribution functions for all hadron species:
\begin{equation}
 f_i(x,p)=f_{i,\text{eq}}(x,p)\left[1+(1\mp f_{i,\text{eq}})\frac{p_\mu p_\nu \pi^{\mu\nu}}{2T^2(\epsilon+p)}\right]
\end{equation}

The scatterings and decays happening after particlization are then treated with UrQMD code.

\section{Results}

First we simulate Pb-Pb collisions at energies $E_\text{lab}=158, 80, 40, 30$ and $20$~A~GeV (i.e.\ $\sqrt{s_{NN}}=17.3,\dots,6.3$~GeV), and compare $m_T$ and $dN/dy$ distributions with the data from NA49 collaboration. Note that the only variable parameter is the shear viscosity to entropy density ratio $\eta/s$ in hydro stage, with the other parameters fixed as described above.

With a given criterion for fluid to particle transition, the duration of hydro phase decreases from about 6.5 fm/c (at $\sqrt{s_{NN}}=17.3$~GeV) to just about 4.5 fm/c (at $\sqrt{s_{NN}}=6.3$~GeV), while the duration of pre-hydrodynamic stage increases according to the Eq. \ref{tstart}. Consequently, at the lowest SPS energy ($\sqrt{s_{NN}}=6.3$~GeV) the lifetimes of pre-hydro (4.1 fm/c) and hydro stages are comparable.

\begin{figure}[h]
 \includegraphics[width=18pc]{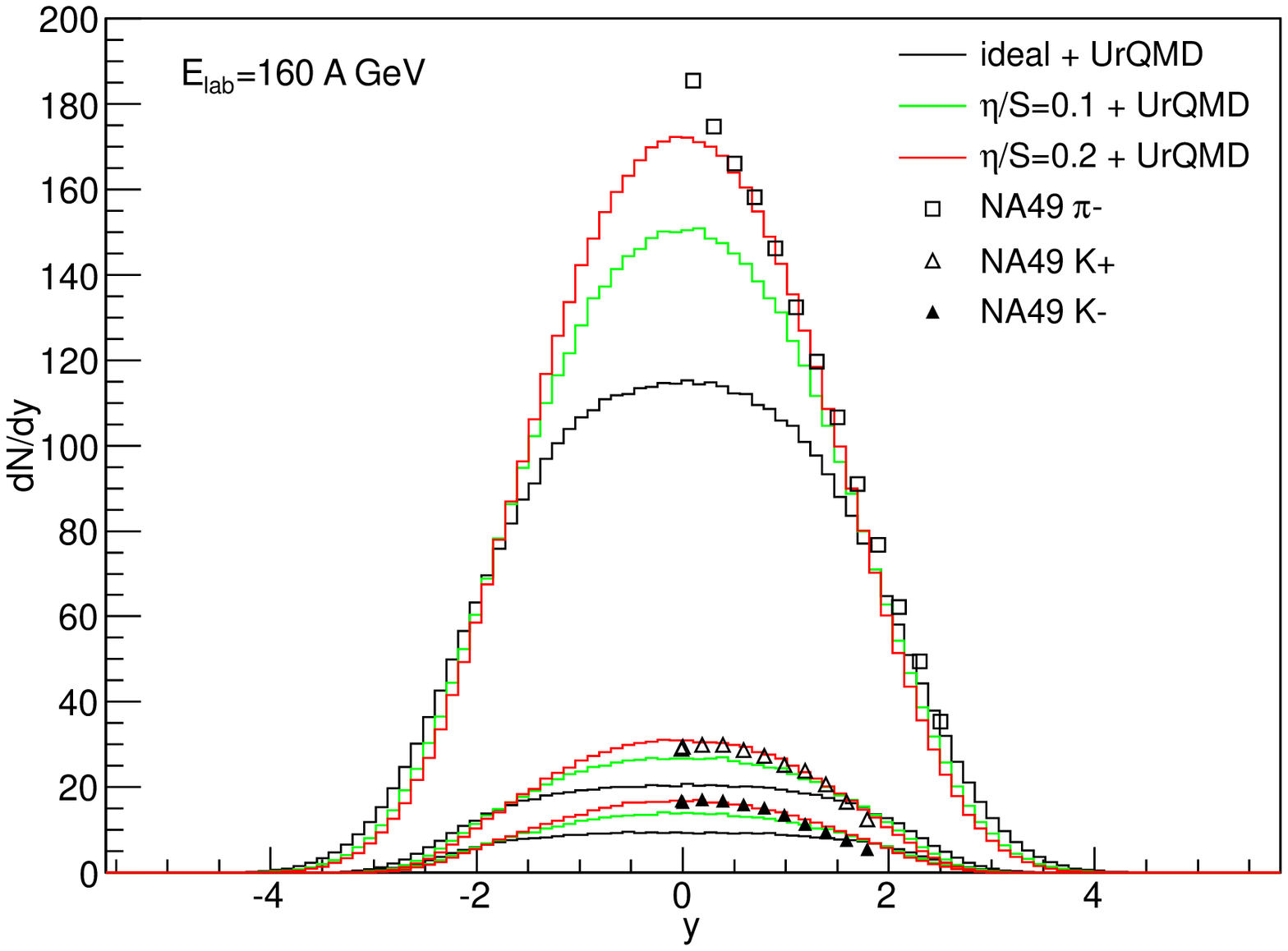}
 \includegraphics[width=17pc]{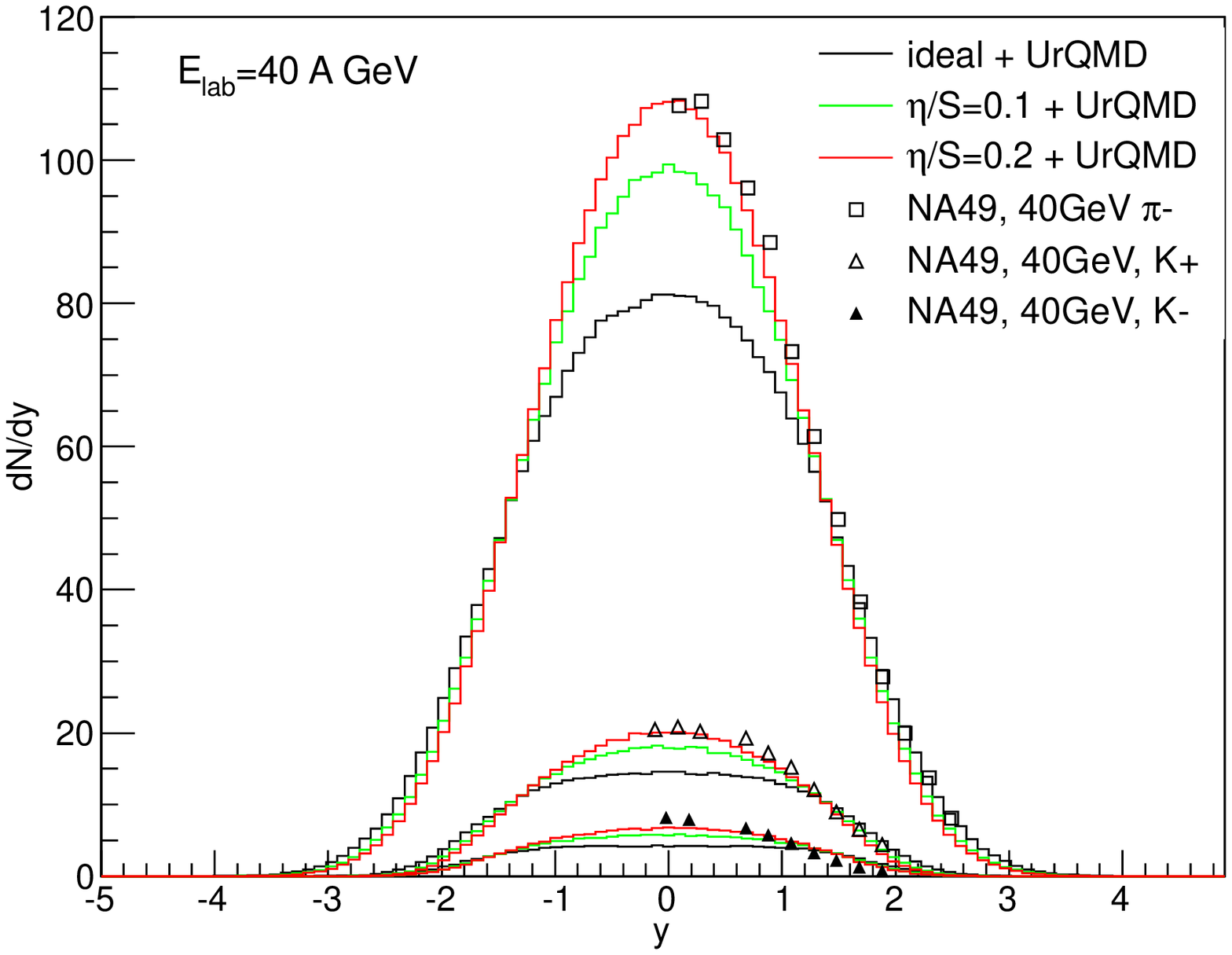}\\
 \caption{\label{figDnDy}Rapidity distributions for $\pi^-$,$K^+$ and $K^-$ for Pb-Pb collisions at $E_\text{lab}=158$ and $40$ A GeV (corresponding to $\sqrt{s_{NN}}=17.3$ and $6.3$ GeV). Model calculations are compared to NA49 data \cite{NA49-40-160}}
\end{figure}

\begin{figure}[h]
 \includegraphics[width=18pc]{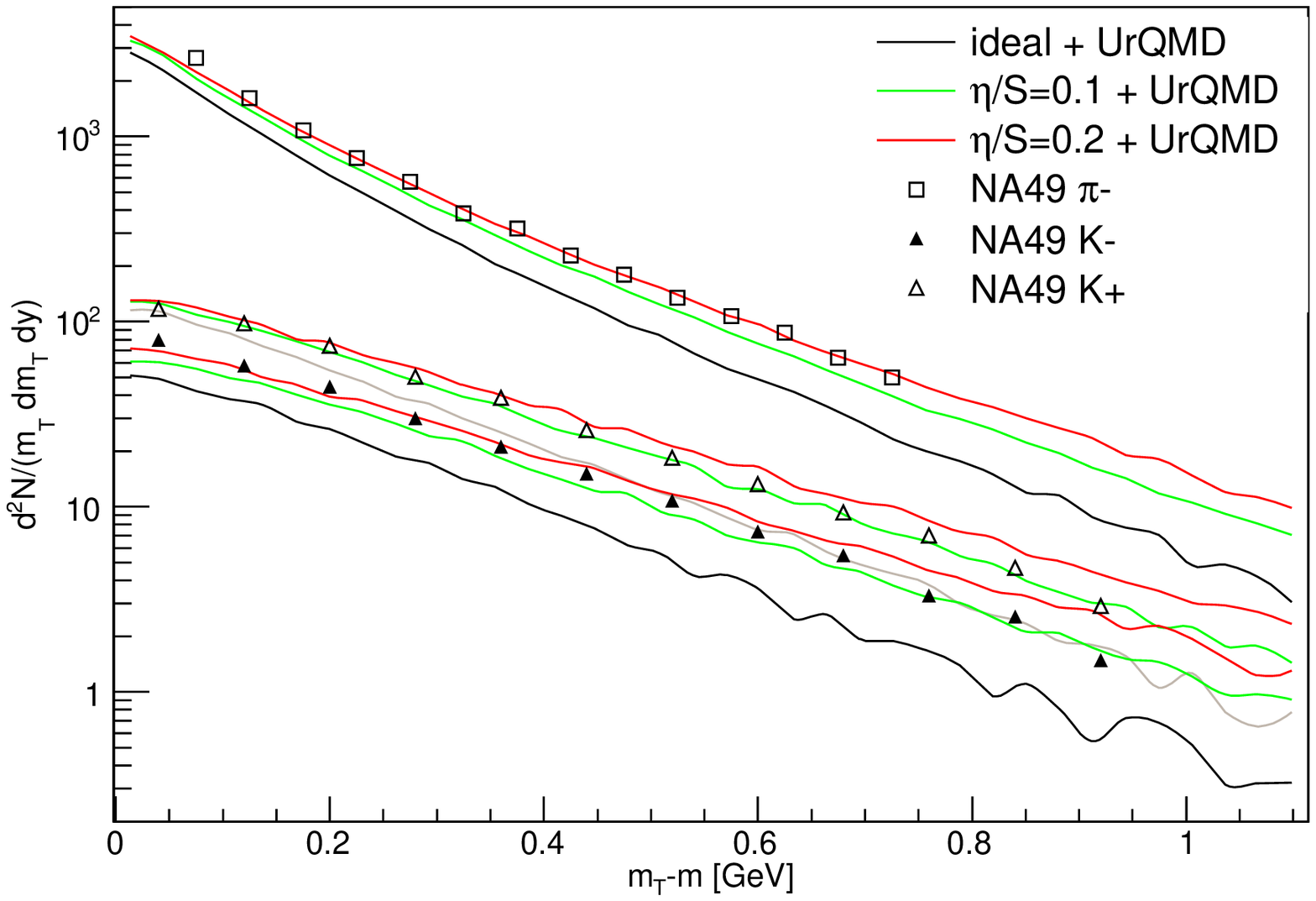}
 \includegraphics[width=18pc]{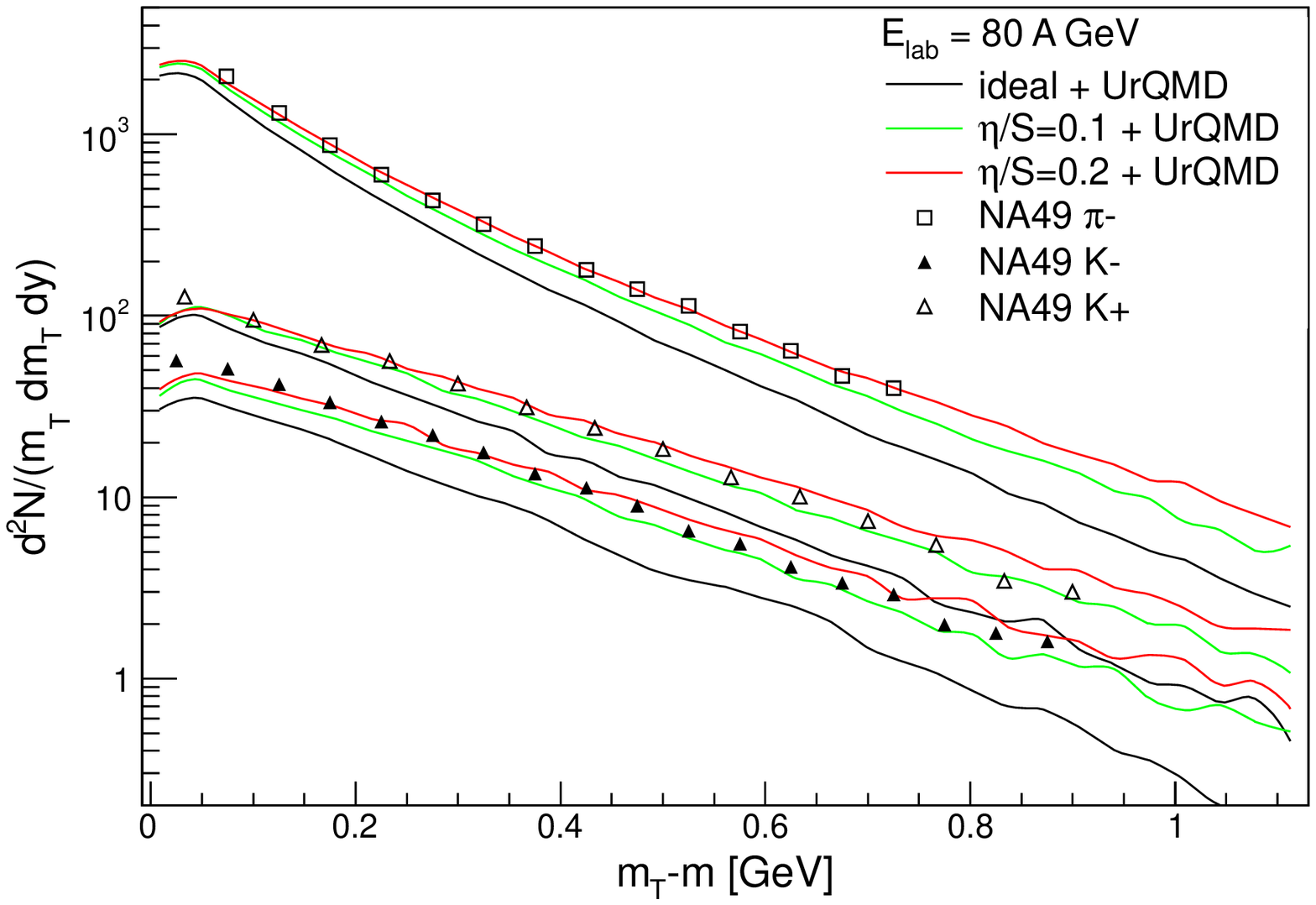}\\
 \includegraphics[width=18pc]{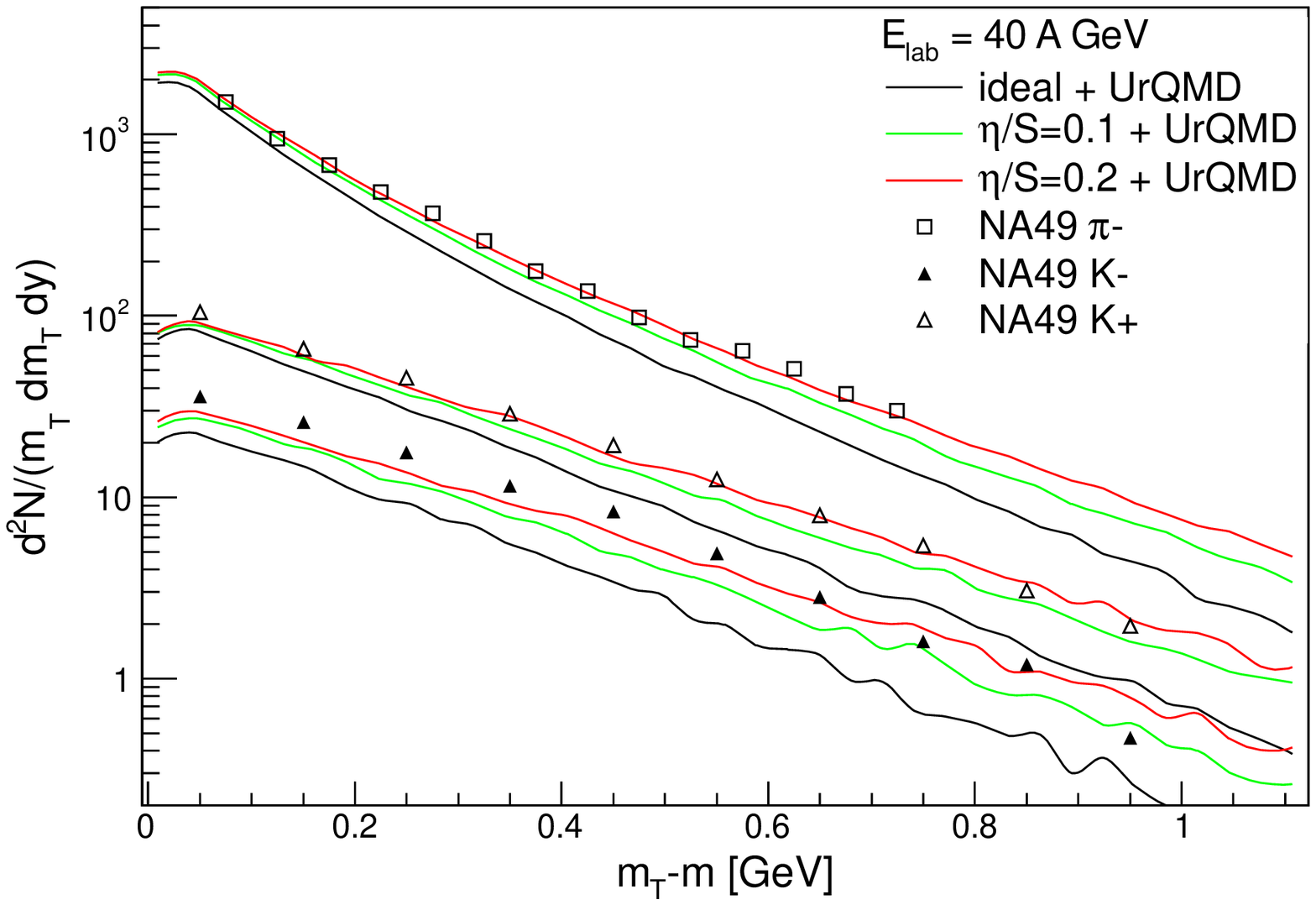}
 \includegraphics[width=18pc]{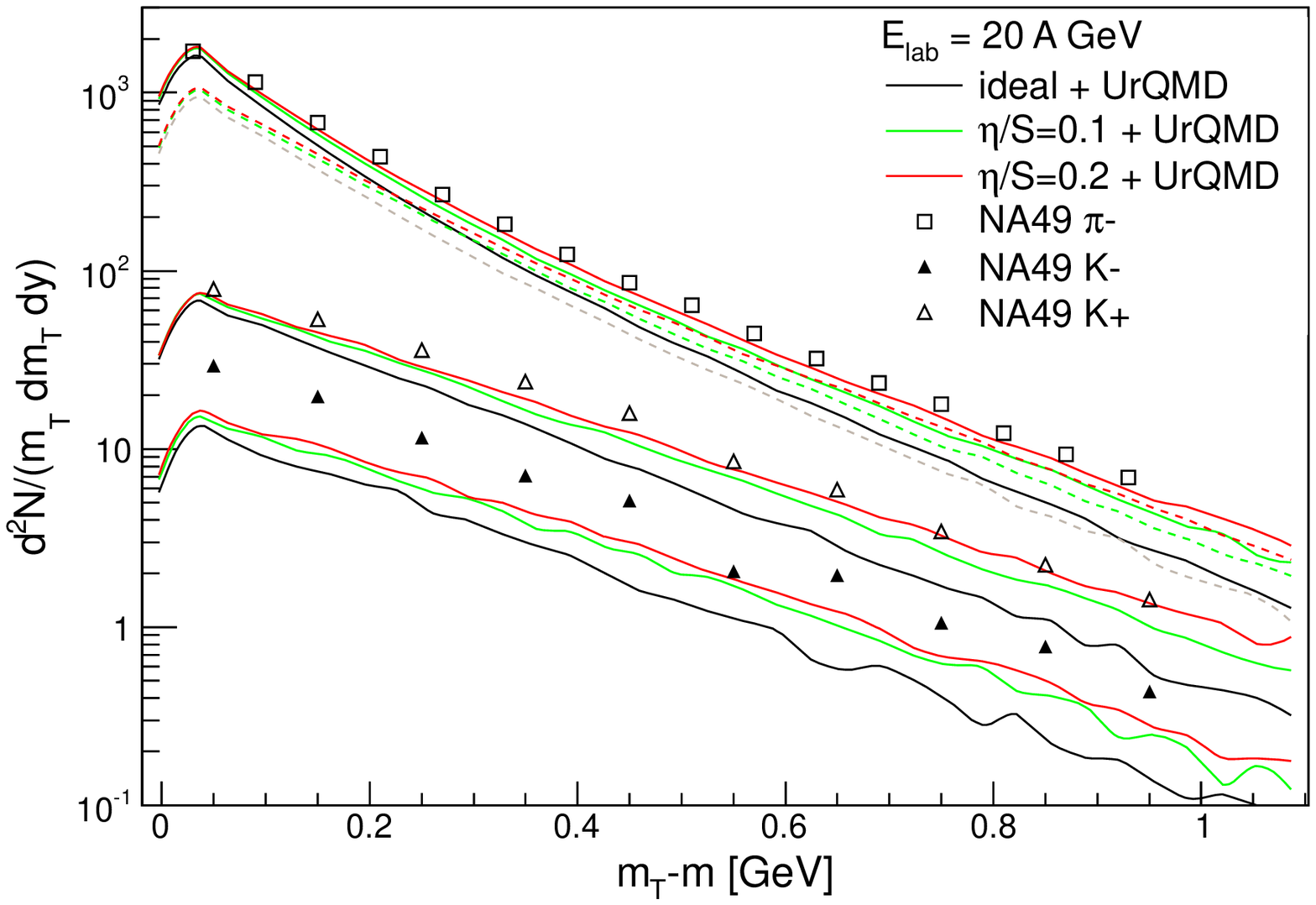}
 \caption{\label{figpt}$m_T$-spectra for $\pi^-$,$K^+$ and $K^-$ for Pb-Pb collisions at $E_\text{lab}=158, 80, 40, 20$~A~GeV ($\sqrt{s_{NN}}=17.3, 12.3, 8.8, 6.3$ GeV, respectively). Model calculations are compared to NA49 data \cite{NA49-20-30, NA49-40-160}. Dashed lines on bottom right plot: model calculations for $\pi^+$.}
\end{figure}

In the case of zero shear viscosity in hydro phase we underestimate the particle yields and radial flow at midrapidity.
However, the inclusion of shear viscosity in hydro phase increases the yields at midrapidity due to viscous entropy production, see Fig. \ref{figDnDy}. One can see that shear viscosity makes the $dN/dy$ profile narrower. The longitudinal expansion is weaker, and the expansion tends to be more spherical, as seen in the comparison of the $p_T$-spectra, Fig. \ref{figpt}: $p_T$-spectra become flatter, which is due to stronger transverse expansion and larger radial flow.

\begin{figure}
 \includegraphics[width=20pc]{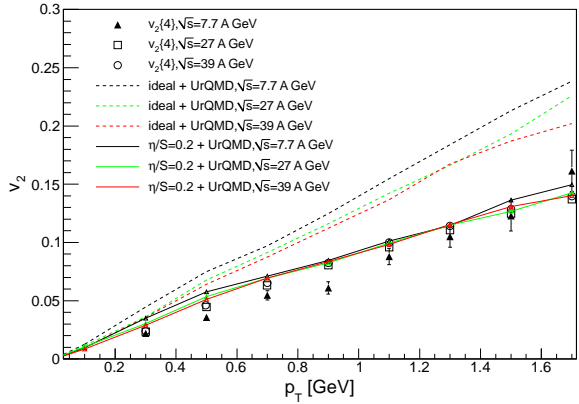}\hspace{2pc}
\begin{minipage}[b]{14pc}\caption{\label{figv2} $p_T$-differential elliptic flow of all charged hadrons for Au-Au collisions at $\sqrt{s_{NN}}=7.7, 27$ and $39$~GeV. The data are from STAR collaboration \cite{starBESv2}}
\end{minipage}
\end{figure}

Next we calculate charged hadron elliptic flow at collision energies $\sqrt{s_{NN}}=7.7, 27$ and $39$~A~GeV to compare with the results from RHIC BES, Fig. \ref{figv2}. Elliptic flow for the case of $\eta/s=0$ overestimates the data, while choosing $\eta/s=0.2$ greatly improves agreement with it. Here we do not try to fit the data and only demonstrate the model results for $\eta/s=0, 0.1$ and $0.2$. So one can conclude that a consistent description of $v_2$, $dN/dy$, and $p_T$-spectra requires a value of $\eta/s$ which is somewhat larger than $0.2$, especially for lower energy points. The $\eta/s=0.2$ also makes $v_2(p_T)$ almost independent of the collision energy.

We conclude that the shear viscosity in hydrodynamic phase is a key component for a hybrid model to better describe the data in the low energy region. The suggested value of the effective shear viscosity is $\eta/s\ge0.2$ for both Pb-Pb collisions at SPS and Au-Au in BES program at RHIC, which is at least twice higher than the typical value of $\eta/s\approx0.08$ \cite{Heinz} obtained for $\sqrt{s_{NN}}=200$~GeV RHIC energy and Monte Carlo Glauber initial state.

\section*{Acknowledgements}
IK and HP acknowledge the financial support by the ExtreMe Matter Institute EMMI and Hessian LOEWE initiative. HP acknowledges funding by the Helmholtz Young Investigator Group VH-NG-822. Computational resources have been provided by the Center for Scientific Computing (CSC) at the Goethe-University of Frankfurt.

\end{document}